\documentclass[usenatbib]{mnras}

\usepackage[T1]{fontenc}

\DeclareRobustCommand{\VAN}[3]{#2}
\let\VANthebibliography\thebibliography
\def\thebibliography{\DeclareRobustCommand{\VAN}[3]{##3}\VANthebibliography}

\usepackage{bm}
\usepackage{graphicx}	

\title[Electron screening correction to the shear modulus]
{Neutron star crust in Voigt approximation II: general formula for electron screening correction for effective shear modulus}

\author[A.\ I. Chugunov]{
	Andrey I. Chugunov,$^{1}$\thanks{E-mail: andr.astro@mail.ioffe.ru}
	\\
	$^{1}$Ioffe Institute, Politekhnicheskaya 26, 194021 St. Petersburg, Russia
}

\date{Accepted XXX. Received YYY; in original form ZZZ}

\pubyear{2022}

\begin{document}
	\label{firstpage}
	\pagerange{\pageref{firstpage}--\pageref{lastpage}}
	\maketitle

\begin{abstract}
	The main contribution to the effective shear modulus of neutron star crust  can be calculated within Coulomb solid model
	 and can be approximated by simple analytical expression
	 for arbitrary (even multicomponent) composition.
	 Here I consider correction associated with electron screening within Thomas-Fermi approximation.   
	In particular, I demonstrate that for relativistic electrons (density $\rho>10^6$~g\,cm$^{-3}$) 
	this correction
	can be estimated as
	$\delta \mu_\mathrm{eff}^\mathrm{V}=
	-9.4\times 10^{-4}\sum_Z n_Z Z^{7/3} e^2/a_\mathrm{e},
	$
	where summation is taken over ion species, $n_Z$ is number density of ions with charge $Ze$, $k_\mathrm{TF}$ is Thomas-Fermi screening wave number. Finally, $a_\mathrm{e}=(4 \pi n_\mathrm{e}/3)^{-1/3}$ is electron sphere radius. Quasineutrality condition $n_\mathrm{e}=\sum_Z Z n_Z$ is assumed.
	This result holds true for arbitrary (even multicomponent and amorphous) matter and can be applied for neutron star crust and (dense) cores of white dwarfs.
    For example, the screening correction reduces shear modulus by $\sim 9$\% for $Z\sim40$, which is typical for inner layers of neutron star crust. 
\end{abstract}



\section{Introduction}

Matter in the neutron star crust as well as  in the cores of white dwarfs solidifies if the temperature becomes low enough (e.g., \citealt*{hpy07,ch08,ch17}). Just like terrestrial solids, solidified stellar matter can support anisotropic stresses. To describe this behaviour, the elasticity theory should be applied. 
The crucial parameter of this theory is the shear modulus, which describes response of the solid to shear deformations. A prescription, which allows to calculate the shear modulus of stellar matter at certain density, composition and temperature is an essential part of the microphysics input required to model a wide range of neutron star phenomenons. These are oscillation spectra, in particular interpretation of the observed quasi-periodic oscillations after giant flares of magnetars (e.g., \citealt{hc80_torsOsc,st83_tors_osc,McDermott_ea88,Strohmayer_etal91,Gabler_ea11,Gabler_ea12,Gabler_ea13,Gabler_ea18,Sotani_ea18,KY20}), the mass distribution asymmetry (mountains) in the crust, which can lead to emission of gravitational waves (see e.g., \citealt{Ushomirsky_ea00,hja06_mountains,Horowitz10_lowmass,McDaniel_Owen13,KM22_Mountain}
for the models and \citealt{LIGO_VIRGO20_MSPelipt,LIGO21_Pulsars,LIGO22_GW_isotated} for recent observational constraints).
Elastisity can also affect tidal deformability,
which leads to quite substantial effects for white dwarf binary evolution (\citealt{Perot_Chamel22}), however for neutron star binaries the effect is almost negligible  \cite{Penner_ea11,Biswas_ea19,Pereira_ea20,Gittins_ea20}.

Neutron star crust and cores of white dwarfs are composed of fully 
ionized atomic nuclei (ions) and degenerate electrons (in the neutron star inner crust unbound neutrons are also present).
It is typically assumed that solidification leads to formation of polycrystalline matter
(e.g., \citealt{oi90,hpy07,ch08} as well as \cite{Caplan_etal18} for molecular dynamics simulations, supporting this assumption)
however, the amorphous state is not fully excluded 
and considered, e.g. by \cite*{Jones99_AmorphousCrust,Jones04_ResistAmorphousCrust,sca20_disorder,cfg20,ah20}.

Elastic properties of stellar matter within polycrystalline assumption were studied in many papers (e.g., \citealt{Strohmayer_etal91,hh08,Baiko11,Baiko12,kp15,Baiko15,Kozhberov19_elast,Kozhberov22}).
Their authors make two steps. First, they assume that ions form a perfect lattice of certain type and apply molecular dynamics or phonon-based formalism to consider response of this system to deformations (note that the first result for static lattice were obtained by \citealt{Fuchs36}).
In this manner the authors calculate elastic constants for perfect crystals, which are anisotropic, i.e. which have more than two independent elastic constants, known for isotropic material: shear and bulk modulus. 
At the second step, anisotropic elastic constants are applied to estimate an effective shear modulus of the polycrystalline matter  $\mu_\mathrm{eff}$ (the bulk modulus is known from the equation of state).
However, as it is well known in terrestrial material science, the second step is far from trivial and, furthermore, $\mu_\mathrm{eff}$ can depend on the correlations in orientations of crystallites (see, e.g., \citealt{ll_elast}, section 10, p.\ 36).
Following \cite{oi90}, astrophysical applications typically apply  Voigt approach (see,  \citealt{kp15} for discussion), which assumes that all crystallites have the same deformation. The Voigt approach gives an upper bound to the effective shear modulus.

The main contribution to the effective shear modulus comes from static lattice, however, corrections associated with ion motion are significant, if temperature is close to the melting point (see e.g., \citealt{hh08,Baiko12}).%
\footnote{Note, that while considering elasticity theory at finite temperature, one should take in mind that the adiabatic and isothermal elastic constants generally differ, however, as shown in \cite{Baiko11}, the Voigt averaged effective shear modulus is the same in both cases (at least for lattices with cubic symmetry).} 
As long as electron screening is rather weak, typically one includes this effect perturbatively within Tomas-Fermi approach, i.e. in the zero-th order one can apply pure Coulomb interaction (Coulomb crystal model)
and include screening effects as a correction.
However, as shown by \cite{Baiko12}, only general trend of the screening correction is well described within Thomas-Fermi approximation, while more accurate treatment on the basis of relativistic electron dielectric function (\citealt{Jancovici62}), leads to complicated nonmonotonic dependence of the effective shear modulus on nuclei proton number $Z$ (see also \citealt{Baiko14,kp21} for complicated zero-temperature phase diagrams of one-component crust, corresponding to \citealt{Jancovici62} dielectric function).

In a recent paper  (\citealt{C21_elastCoins}; paper I below), I demonstrate that 
the Voigt-averaged effective shear modulus can be calculated analytically for arbitrary (even multicomponent or/and amorphous) solid, if ion vibrations are neglected and Coulomb crystal model is applied. 
Estimating Madelung energy within ion sphere model (ISM), I obtain $\mu^\mathrm{C}_\mathrm{eff}\approx\sum_Z 0.12\, n_Z Z^{5/3}e^2 /a_\mathrm{e}$. Here summation is taken over ion species, $n_Z$ is number density of ions with charge $Ze$. Finally, $a_\mathrm{e}=(4 \pi n_\mathrm{e}/3)^{-1/3}$ is electron sphere radius. Quasineutrality condition $n_\mathrm{e}=\sum_Z Z n_Z$ is assumed. 

Motivated by numerical results by \cite{Kozhberov22}, in this letter I expand  Paper I by including electron screening correction within Tomas-Fermi approach, still neglecting ion vibrations (e.g.\ assuming zero temperature and neglecting zero-point motion).
In particular, I demonstrate that electron screening decreases the effective shear modulus by the amount $\delta^\mathrm{scr} \mu_\mathrm{eff}=(4/15) k_\mathrm{TF}^2 a_\mathrm{e}^2 \epsilon_\mathrm{scr}$,
where $\epsilon_\mathrm{scr}<0$ describes screening correction to the energy density.%
\footnote{Similar result were obtained recently by \cite{Khrapak19} for the transverse sound velocities of Yukawa systems, but his derivation is based on quasicrystalline approximation, initially suggested for liquids by \cite{Hubbard_1969}, so the details of averaging of anisotropic elastic tensors at finite pressure remain unclear.}
Here 
\begin{equation}
k_\mathrm{TF}^2=4\pi e^2\frac{\partial n_\mathrm{e}}{\partial \mu_\mathrm e}\approx
0.0342\frac{\sqrt{1+x_\mathrm{r}^2}}{x_\mathrm{r}}
\label{kTF}
\end{equation}
is Thomas-Fermi screening wave number, $\mu_\mathrm{e}=m_\mathrm{e} c^2\sqrt{1+x_\mathrm{r}^2}$ is  chemical potential of electrons, $x_\mathrm{r}\equiv \hbar (3\pi^2 n_\mathrm{e})^{1/3}/(m_\mathrm{e} c)$ is electron relativity parameter.
Finally,  $m_\mathrm{e}$, $c$, and $\hbar$ are electron mass, speed of light, and the Planck constant.
 
Estimating $\epsilon_\mathrm{scr} $ via ISM (\citealt{Khrapak_ea14_ISM}),  one can write the screening correction  as
\begin{equation}
\delta \mu_\mathrm{eff}^\mathrm{V}=
-0.027 k_\mathrm{TF}^2 a_\mathrm{e}^2\sum_Z n_Z \frac{Z^{7/3} e^2}{a_\mathrm{e}}.
\end{equation}
As in paper I, this result holds true for arbitrary composition and microphysical structure of solid.

This letter generally follows the structure of the paper I: a brief introduction to the finite pressure elasticity theory is given in section \ref{Sec_ElastTheor} (see, in particular,  a methodical note, section \ref{Sec:methodnote});
approximations applied in the letter are formulated in section \ref{Sec:Assum};
section \ref{Sec:Sym} demonstrates symmetry of elasticity tensor within Thomas-Fermi approximation and uses this symmetry to derive the effective shear modulus.
Short summary is presented in section \ref{Sec:summary}.

\section{Elastisity tensors of stellar solids: screening correction}

\subsection{Elastisity at finite pressure}\label{Sec_ElastTheor}
To introduce notations and to remind about basic features, I start from a short introduction to the elasticity theory in application to stellar solids.

The elasticity theory describes response of the matter to deformation, i.e.\ displacement of the matter element from the point $\bm{R}$ to $\tilde{\bm{R}}=\bm{R}+{\bm \xi}(\bm{R})$, where ${\bm \xi}(\bm{R})$ is the displacement vector. 
I will consider uniform infinitesimal deformations:
$\xi(\bm R)_i=u_{ij}R_j$, where $u_{ij}$ is 
displacement gradient, which is assumed to be constant over 
solid. 

For astrophysical applications one should consider deformation with respect to the initial state, which has finite pressure $P$. It makes elasticity theory a bit more complicated than the standard textbook version (e.g., \citealt{ll_elast}). In particular, several elasticity tensors should be introduced.
Following \cite{Wallace67}, let us start from the general form for the change of the energy [per unit volume at initial (undeformed) state], written up to second order terms in $u_{ij}$:
\begin{equation}
\delta E=\sigma_{ij} u_{ij}+\frac{1}{2} S_{ijkl} u_{ij} u_{kl}. \label{S_def}
\end{equation}
Here $\sigma_{ij}$ is the stress tensor at initial state, assumed to be isotropic 
($\sigma_{ij}=-P\delta_{ij}$, where $\delta_{ij}$ is Kronecker delta).
Eq.\ (\ref{S_def}) can be treated as definition of the tensor $S_{ijkl}$, which
does not have Voigt symmetry 
(for example, $S_{ijkl}\ne S_{jikl}$ if $i=k$, $j=l$, $k\ne l$, and $P\ne0$).
Tensor $S_{ijkl}$ is useful for first principle calculation of the elastic properties and will be used in derivations below.

However, for astrophysical applications the stress-strain tensor $B_{ijkl}=S_{ijkl}-
P\left(\delta_{il}\delta_{jk}
-\delta_{ij}\delta_{kl}\right)$ seems to be more important because it allows to calculate the change of the stress tensor $\delta \sigma_{ij}$, associated with the deformation
\begin{equation}
\delta \sigma_{ij}=\frac{1}{2} B_{ijkl}\left(u_{kl}+u_{lk}\right).
\label{B_def}
\end{equation}
It is easy to check that 
$B_{ijkl}+B_{ilkj}=S_{ijkl}+S_{ilkj}$. 
As shown by \cite{Wallace67},  tensor $B_{ijkl}$ has Voigt symmetry
($B_{ijkl}=B_{jikl}=B_{ijlk}=B_{klij}$), and, thus, up to 21 independent elastic parameters.

For isotropic material, the stress-strain tensor $B_{ijkl}$ has only two independent elastic constants $K$ and $\mu$, which are bulk and the shear moduli, respectively:
\begin{equation}
B_{ijkl}=K\delta_{ij}\delta_{kl}
+\mu\left(\delta_{ik}\delta_{jl}+\delta_{il}\delta_{jk}
-\frac{2}{3}\delta_{ij}\delta_{kl}\right).
\label{BV}
\end{equation}
Respective  stress-strain relation is well known
\begin{eqnarray}
\delta \sigma_{ij}
=K\delta_{ij} u_{ll}
+\mu\left(u_{ij}+u_{ji}-\frac{2}{3}\delta_{ij}u_{ll}\right).
\label{delta_sigma_isotrop}
\end{eqnarray}

In the case of  stellar matter, the energy (and thus $S_{ijkl}$) can be presented as a sum of partial contributions from degenerate electrons and electrostatic interaction of ions (for inner crust, contribution of unbound neutrons should be added). 
In this letter I will apply  Voigt average,
which is based on two assumptions: 1) crystallites of polycrystalline matter 
	are randomly oriented and 2) $u_{ij}$ is the same for all crystallites.
This approach is linear over $S_{ijkl}$ and leads to isotropic form of the stress-strain tensor.
Below, the effective shear modulus, obtained by Voigt average is marked as $\mu_\mathrm{eff}^\mathrm{V}$, while the bulk modulus is denoted as $K^\mathrm{V}$.
Only  electrostatic interaction contributes to $\mu_\mathrm{eff}^\mathrm{V}$ and  in what follows $S_{ijkl}$ represents only this part of respective tensor.

\subsubsection{Methodical note}\label{Sec:methodnote}

The difference between tensors $B_{ijkl}$ and $S_{ijkl}$ has simple physical nature: it is associated with compression of the matter element at the second order in $u_{ij}$.
Namely, the energy change of the matter element is equal to the work performed by exterior forces over this solid. In case of finite pressure at initial state, this work has a component $-P\delta V $, where $\delta V$ is the volume change. This term should be calculated up to the second order in $u_{ij}$ to produce correct form of Eq.\ (\ref{S_def}):
\begin{equation}
\delta V=V \left(\delta_{ij} u_{ij}+\frac{\delta_{ij}\delta_{kl}-\delta_{il}\delta_{jk}}{2} u_{ij}u_{kl}
\right),
\label{dV}
\end{equation}
The first term in the brackets is well known first-order relative change of the volume, while the second is the second-order change of the volume, which both contribute to $\delta E$ by the term $\propto P$. It is  work associated with this term that leads to the difference between $S_{ijkl}$ and $B_{ijkl}$ tensors.

Alternative formulation of the finite pressure elasticity  theory  were suggested by \cite{Marcus_ea02,Marcus_Qiu04_reply,Marcus_Qui09}. It is based on the expansion of  $G=E+PV-TS$ ($T$ and $S$ are the temperature and the entropy, in this paper I consider $T=0$ limit). Indeed, thanks to the fact that variation of $G$ is considered, the term $\propto \delta V$ is cancelled out and there are no need to introduce tensor $S_{ijkl}$ (change of the $G$ is described by the tensor $B_{ijkl}$).
In the papers by \cite{Marcus_ea02,Marcus_Qiu04_reply,Marcus_Qui09}  $G$ is refereed  as the Gibbs energy, but it can be misleading (see, e.g., \citealt{Steinle_Neumann_2004_Marcus_crit}).
Namely, within $G$-based approach by \cite{Marcus_ea02} the quantity $P$  is treated as a parameter, which is
not modified by deformation (even if one considers compression, which leads to change of the pressure). As a result,  $G=E+PV-TS$ has a minimum, corresponding to equilibrium state at the pressure, which is equal to $P$ 
(see \citealt{Marcus_Qiu04_reply} for detailed clarification and arguments that $G$ is a natural choice for some problems, e.g., for sound waves in matter under fixed external pressure).
However, applying the theory by \cite{Marcus_ea02}, one should keep in mind that $G$ is not uniquely determined for given microscopic state (e.g., position of all ions in section \ref{Sec:Sym}) because it depends on the value of the parameter $P$, i.e. on the state of the matter with respect to which deformations are considered. 
As a result, $G$ generally  cannot be treated as thermodynamic property and thus should be distinguished from the Gibbs energy as it is determined by equation of state (microscopic calculations at isotropic stress).
That is why I prefer to appeal to the finite pressure elasticity theory as it is formulated by  \cite{Wallace67} (if section \ref{Sec:Sym} were written within \citealt{Marcus_ea02} approach, the second order terms would become more complicated).

Hopefully, in astrophysical applications one is typically interested in variation of the stress tensor, given by Eq.\ (\ref{B_def}) (often in isotropic form, Eq.\ \ref{delta_sigma_isotrop}) and the above mentioned complications are not important, if properly calculated elastic constants ($\mu$ and $K$ for Eq.\ \ref{delta_sigma_isotrop}) are applied.

In this work I present
screening correction to $\mu^\mathrm{V}_\mathrm{eff}$ (pure Coulomb result was presented in Paper I). Screening correction to the bulk modulus, $K=-V\left(\mathrm d P/\mathrm d V \right)$,
can be obtained from the equation of state, which is well known (e.g., \citealt{kp21} and references therein).

\subsection{Physical model: approximations}
\label{Sec:Assum}
I consider screening correction to the elastic properties of neutron star crust and white dwarf cores within the following widely applied approximations:
\begin{enumerate}
\item All ions are fully ionized and  considered as point charges;
\item Ions are static (i.e.\ located at equilibrium positions, thermal and zero point vibrations are neglected);
\item The deformation is uniform at microphysical level (e.g. $u_{ij}=\mathrm{const}$);
\item Electron screening is described within linear response theory with static longitudinal dielectric function, corresponding to the Thomas-Fermi approximation:
$	\epsilon(q)=1+k_\mathrm{TF}^2/q^2$.
\end{enumerate}

In this paper I am mostly interested in incompressible  shear deformations, which do not affect the dielectric function. To simplify discussion in the next section I will refer to the Yukawa system, i.e. system of point-like particles interacting via an effective Yukawa potential with (fixed) screening parameter $\kappa=k_\mathrm{TF}$.
For incompressible shear deformations this system is equivalent to the stellar matter (within above-mentioned assumptions), while for deformations with compression it allows to simplify derivations (brute force consideration should take into account that compression affects $k_\mathrm{TF}$). Thanks to linearity of the Voigt averaging, this simplification does not affect $\mu_\mathrm{eff}^\mathrm V$.
However,  to avoid confusions, let me stress that for astrophysical applications the bulk modulus $K$  should be calculated from the equation of state, but not from the formulae of section \ref{Sec:Sym}, which neglect variation of $k_\mathrm{TF}$ with compression and, furthermore, do not have the dominating contribution from degenerate electrons.

\subsection{Symmetry of the elasticity tensor for stellar matter 
and effective shear modulus}
\label{Sec:Sym}

Following Paper I, to derive 
$S_{ijkl}$ tensor, I calculate a change of the electrostatic energy $\Delta E$, associated with deformation. 
Within applied approximations (section \ref{Sec:Assum}), electrostatic energy of the matter element
can be written as
\begin{eqnarray}
E&=& 
\frac{1}{2}\sum_{a} \sum_{b\ne a}
\frac{Z^a Z^b e^2}{\left|\bm R^{a}-\bm R^{b}\right|}
\exp\left(-\kappa \left|\bm R^{a}-\bm R^{b}\right| \right)
\nonumber \\
&-&  \sum_a\int \frac{Z^a e^2 n_\mathrm{e} }{\left|\bm R^{a}-\bm r\right|}
\exp\left(-\kappa \left|\bm R^{a}-\bm r\right| \right)
\mathrm d^3 \bm r
\label{energy} \\
&+&\frac{1}{2}
\int\int  \frac{e^2 n_\mathrm{e}^2}{\left|\bm r- \bm r^\prime\right|} 
\exp\left(-\kappa \left|\bm r^\prime-\bm r\right| \right)
\mathrm d^3 \bm r\,
\mathrm d^3 \bm r^\prime.
\nonumber  
\\
&-&\frac{\kappa}{2}\sum_{a} Z_a^2.
\nonumber
\end{eqnarray}
The 
first three 
terms in equation (\ref{energy})
are ion-ion, ion-electron, and electron-electron interaction energies respectively.
The last term corresponds to the electrostatic interaction energy of each nuclei with the  screening cloud; it is responsible for the enhancement of nuclear reaction rate by electron screening (e.g., \citealt{Salpeter54}).
Here upper indices $a$ and $b$ enumerate ions, $Z^a e$ and $\bm R^a$ are the charge and the position  of ion $a$ respectively.
The electron number density is 
 $n_\mathrm{e}=\sum_a Z^a/V$ due to quasineutrality condition. 

After applying deformation $u_{ij}$, the energy becomes $\tilde E$ and the ion positions become
\begin{equation}
\tilde{R}_i^a=R_i^a+u_{ij}R^a_j.
\label{tildeRa}
\end{equation}
Taking into account that for electrons this deformation modifies the number density (it becomes $\tilde n_e =n_e/J$, where $J$ is the Jacobian of transformation; the quasineutrality condition obviously holds true after deformation), the energy
change 
$\Delta E=\tilde E-E$
 can be written in the form
\begin{eqnarray}
\delta E&=& \tilde E-E=
\frac{1}{2}\sum_{a} \sum_{b\ne a}
Z^a Z^b e^2
\left(
 \frac{\mathrm e^{-\kappa  \left|\tilde {\bm \Delta}^{ab} \right|}}
 {\left|\tilde {\bm \Delta}^{ab} \right| }  
 -\frac{\mathrm e^{-\kappa  \left| {\bm \Delta}^{ab} \right|}}{\left| {\bm \Delta}^{ab} \right| }
\right)
\nonumber \\
&-& 
\sum_a\int Z_i e^2 n_\mathrm{e} 
\left(
  \frac{\mathrm e^{-\kappa\left|\tilde {\bm \Delta}^{ar} \right|}}{\left|\tilde {\bm \Delta}^{ar} \right|  }  
 -\frac{\mathrm e^{-\kappa \left| \bm \Delta^{ar} \right|}}{\left| \bm \Delta^{ar} \right| } 
\right)\mathrm d^3 \bm r
\nonumber \\
&+&\frac{1}{2}
\int\int  e^2 n_\mathrm{e}^2
 \left(\frac{\mathrm e^{-\kappa \left|\tilde {\bm \Delta}^{rr} \right|}}{\left|\tilde {\bm \Delta}^{rr} \right|} 
  -\frac{\mathrm e^{-\kappa\left| \bm \Delta^{rr} \right|}}{\left| \bm \Delta^{rr} \right|} 
  	 \right)
\mathrm d^3 \bm r\,
\mathrm d^3 \bm r^\prime.
\label{denergy}
\end{eqnarray}
Here, $\bm \Delta^{ab}=\bm R^{a}-\bm R^{b}$, $\bm \Delta^{ar}=\bm R^{a}-\bm r$,  and  $\bm \Delta^{rr}=\bm r-\bm r^\prime$. Similarly, for the deformed state: $\tilde {\bm \Delta}^{ab}=\tilde{\bm R}^{a}-\tilde{\bm R}^{b}$,
$\tilde {\bm \Delta}^{ar}=\tilde{\bm R}^{a}-\tilde{\bm r}$, and
$\tilde {\bm \Delta}^{rr}=\tilde{\bm r}-\tilde{\bm r^\prime}$.

Applying Taylor expansion over  
$u_{ij}$, taking into account that $\tilde {\Delta}^\alpha_i=\Delta^\alpha_i+u_{ij}\Delta^\alpha_j$ (here and below index $\alpha$ runs over $ab$, $ar$, and $rr$):
\begin{eqnarray}
\frac{\mathrm e^{-\kappa \left| \tilde{\bm \Delta}^{\alpha} \right|}}{\left|\tilde {\bm \Delta}^{\alpha} \right|}
&\approx&
\frac{\mathrm e^{-\kappa \left| {\bm \Delta}^{\alpha} \right|}}{ \left| {\bm \Delta}^{\alpha} \right| }
+p^\alpha_{ij} u_{ij}
+s^\alpha_{ijkl}u_{ij}u_{kl},
\label{Taylor}
\end{eqnarray}
where
\begin{eqnarray}
 p^\alpha_{ij}&=&-\frac{\mathrm e^{-\kappa \left| {\bm \Delta}^{\alpha} \right|}}
 {\left| {\bm \Delta}^{\alpha} \right|^3}\Delta^\alpha_i \Delta^\alpha_j
 \left(\kappa\left| {\bm \Delta}^{\alpha} \right|+1\right)\\
s^\alpha_{ijkl}&=&\frac{\mathrm e^{-\kappa \left| {\bm \Delta}^{\alpha} \right|}}
{\left| {\bm \Delta}^{\alpha} \right|^5}\Delta^\alpha_j \Delta^\alpha_l 
\left\{\Delta^\alpha_i \Delta^\alpha_k\left[3 \kappa \left| {\bm \Delta}^{\alpha} \right|+\kappa^2\left| {\bm \Delta}^{\alpha} \right|^2+3\right] \right.
\nonumber \\
&&\left. -\left| {\bm \Delta}^{\alpha} \right|^2\left(\kappa\left| {\bm \Delta}^{\alpha} \right|+1\right) \delta_{ik}\right\}.
\end{eqnarray}
Combining Eqs.\ (\ref{Taylor}) and (\ref{denergy}), one can identify that 
\begin{eqnarray}
V\sigma_{ij}&=&\frac{1}{2}\sum_{a} \sum_{b\ne a}
Z^a Z^b e^2 p_{ij}^{ab}
-\sum_a\int Z^a e^2 n_\mathrm{e} p_{ij}^{ar}
\mathrm d^3 \bm r
\nonumber \\
&+&\frac{1}{2}
\int\int  e^2 n_\mathrm{e}^2 p_{ij}^{rr}
\mathrm d^3 \bm r\,
\mathrm d^3 \bm r^\prime,
\label{sig_ij} 
\end{eqnarray}
while 
\begin{eqnarray}
VS_{ijkl}&=&\frac{1}{2}\sum_{a} \sum_{b\ne a}
Z^a Z^b e^2 s_{ijkl}^{ab}
-\sum_a\int Z^a e^2 n_\mathrm{e} s_{ijkl}^{ar}
\mathrm d^3 \bm r
\nonumber \\
&+&\frac{1}{2}
\int\int  e^2 n_\mathrm{e}^2 s_{ijkl}^{rr}
\mathrm d^3 \bm r\,
\mathrm d^3 \bm r^\prime
\label{Sijkl} 
\end{eqnarray}

To calculate contractions $\sigma_{ii}$ and  $S_{ijij}$ let me note that 
\begin{eqnarray}
p^\alpha_{ii}&=&-\frac{\kappa \left| {\bm \Delta}^{\alpha} \right|+1}{\left| {\bm \Delta}^{\alpha} \right|}\mathrm e^{-\kappa \left| {\bm \Delta}^{\alpha} \right|},\\
s^\alpha_{ijij}&=&\kappa^2\left| {\bm \Delta}^{\alpha} \right| \mathrm e^{-\kappa \left| {\bm \Delta}^{\alpha} \right|}.
\end{eqnarray}

Now it is easy to compare contractions $\sigma_{ii}$ and  $S_{ijij}$ with derivatives of $E$  over $\kappa$.
Since each sum and integral in Eq.\ (\ref{energy}) converge due to the factor $\exp\left(-\kappa \Delta^\alpha\right)$, the derivative can be taken inside the integral/sum.
It leads to
\begin{eqnarray}
\sigma_{ii}&=&\frac{\kappa}{V}\frac{\partial E}{\partial \kappa} - \frac{E}{V}, \label{deriv_sigm}\\
S_{ijij}&=&\frac{\kappa^2}{V}\frac{\partial^2 E}{\partial \kappa^2}. \label{deriv_S}
\end{eqnarray}
According to \cite{Baiko02,kp21}, in the lowest order in $\kappa$, the energy density $\epsilon=E/V=\epsilon_\mathrm{C}+\kappa^2a_\mathrm{e}^2 \epsilon_\mathrm{scr}$, where $\epsilon_\mathrm{C}$ is energy density in the absence of screening and $\epsilon_\mathrm{scr}$ describes screening correction. Thus,
\begin{eqnarray}
 \sigma_{ii}&=&-\epsilon_\mathrm{C}+\kappa^2 a_\mathrm{e}^2\epsilon_\mathrm{scr}, \label{res_sigma}\\
S_{ijij}&=&2\kappa^2 a_\mathrm{e}^2\epsilon_\mathrm{scr} \label{res_S}. 
\end{eqnarray}

As follows from (\ref{res_sigma}), the electrostatic contribution to the pressure is
\begin{equation}
P=-\frac{1}{3}\sigma_{ii}=\frac{1}{3}\epsilon_C-\frac{1}{3}\kappa^2 a_\mathrm{e}^2\epsilon_\mathrm{scr}.
\end{equation}
It agrees with zero-temperature thermodynamic relation $P=-\partial E /\partial V$, taking into account that $\epsilon_\mathrm{C}$ and $\epsilon_\mathrm{scr}$ are both $\propto n_\mathrm{e}^{4/3}$ (see \citealt{Baiko02,kp21} and Eqs.\ \ref{e_c_ISM} and \ref{e_scr_ISM}; as pointed in Section \ref{Sec:Assum}, $\kappa$ treated as  a constant here).

Similar to paper I, Equation (\ref{res_S}) allows to calculate screening correction to the Voigt averaged effective shear modulus $\mu_\mathrm{eff}^\mathrm{V}$. Let us note two facts (see, e.g., Paper I): 1) the contraction $S_{ijij}=B_{ijij}$; 2) $S_{ijij}$ is invariant with respect to Voigt average. As a result, (\ref{res_S}) should hold true for $B_{ijij}^\mathrm{V}$, where $B_{ijkl}^\mathrm{V}$ is the Voigt averaged stress-strain tensor. It is isotropic and thus can be written in form (\ref{BV}). Straightforward calculations lead to 
\begin{equation}
B_{ijij}^\mathrm{V}=3\,K_\mathrm{eff}^\mathrm{V}+10\mu_\mathrm{eff}^\mathrm{V}.\label{B_ijij}
\end{equation}
Considering isotropic compression, I conclude 
\begin{equation}
K_\mathrm{eff}^\mathrm{V}=n_\mathrm{e}\frac{\partial P}{\partial n_\mathrm{e}}=
\frac{4}{9}\epsilon_\mathrm{C}-\frac{2}{9}\kappa^2 a_\mathrm{e}^2\epsilon_\mathrm{scr}.
\end{equation}
Inserting $K_\mathrm{eff}^\mathrm{V}$ into (\ref{res_S}) and (\ref{B_ijij}) leads to conclusion
\begin{equation}
\mu_\mathrm{eff}^\mathrm{V}=-\frac{2}{15}\epsilon_\mathrm{C}+\frac{4}{15} \kappa^2 a_\mathrm{e}^2 \epsilon_\mathrm{scr},
\label{res}
\end{equation}
which is the main result of the paper (the first term was already obtained in Paper I).

Quantities  $\epsilon_\mathrm{C}$ and $\epsilon_\mathrm{scr}$ can be 
precisely calculated, if  lattice structure is given (e.g., \citealt{Baiko02,kb15,cf16_mix,Kozhberov18_Yuk,kp21}). However, taking into account approximate nature of the Voigt average,
it is reasonable to
apply
ISM to estimate energy density (\citealt{Khrapak_ea14_ISM}) :
\begin{eqnarray}
\epsilon^\mathrm{ISM}_\mathrm{C}\approx-0.9 \sum_Z n_Z \frac{Z^{5/3} e^2}{a_\mathrm{e}},
\label{e_c_ISM}\\
\epsilon^\mathrm{ISM}_\mathrm{scr}\approx-0.103 \sum_Z n_Z \frac{Z^{7/3} e^2}{a_\mathrm{e}}.
\label{e_scr_ISM}
\end{eqnarray}
Using Eq.\ (\ref{kTF}), the final estimate is
\begin{eqnarray}
\mu_\mathrm{eff, ISM}^\mathrm{V}
&\approx&
0.12\sum_Z n_Z  \frac{Z^{5/3} e^2}{a_\mathrm{e}}
\label{res_ISM}
 \\
&-&9.4\times 10^{-4}\frac{\sqrt{1+x_\mathrm{r}^2}}{x_\mathrm{r}} \sum_Z n_Z \frac{Z^{7/3} e^2}{a_\mathrm{e}}.
\nonumber
\end{eqnarray}

It is worth stressing, that ISM is surprisingly accurate.
For example, instead of $0.9$  in 
Eq.\ (\ref{e_c_ISM}) precise calculations for body centered cubic (bcc) and face centered cubic (fcc) lattices by \citealt{Baiko02} give $0.895929255682$ and $0.895873615195$, while for screening correction, instead of $0.103$ in (\ref{e_scr_ISM}) accurate calculations give coefficients $0.103732333707$ and $0.103795687531$ for bcc and fcc respectively.

As pointed in the introduction, Eq.\ (\ref{res}) were previously obtained numerically by \cite{Kozhberov22} for bcc and fcc lattices.
Similar result were also obtained by \cite{Khrapak19} for the sound velocities of Yukawa systems.

\section{Summary}\label{Sec:summary}
In this paper I demonstrate that the screening correction to the Voigt averaged shear modulus of neutron star crust and white dwarf cores can be easily calculated via Eq.\ (\ref{res}), if the screening correction to the electrostatic energy are known (in Paper I similar result were obtained for the main Coulomb term).
Estimating the latter within ISM, which is known to be extremely accurate for bcc and fcc lattices,  allows to obtain explicit numerical formula (\ref{res_ISM}).
This result is applicable for arbitrary composition (even multicomponent) and structure (crystalline or amorphous) and can be directly applied in the astrophysics.
For example, for ions with $Z=40$, which are typical for  inner layers of neutron star crust (e.g., \citealt{DH01,Pearson_ea18,Pearson_ea19_erratum,Carreau_ea20_inner}), the screening reduces the shear modulus by $\sim 9$\%. It can be important, e.g.\ for torsional oscillations of neutron star
crust, with frequency $\propto \sqrt \mu$ (e.g., \citealt{KY20}).

However, several points are important.
First, the Voigt average gives an upper limit for the shear modulus of polycrystalline matter and actual shear modulus can be lower (see, e.g., \citealt{kp15} and Paper I). 
Second, consideration in this work neglects vibration of ions, which reduces the shear modulus, especially near the melting point (e.g., \citealt{Baiko12}).
Thus it is better to treat  (\ref{res_ISM}) as an upper limit rather than an exact shear modulus of stellar matter.

\section*{Acknowledgements}
I'm grateful to A.A.~Kozhberov, who provided me an unpublished (at that time) version of  \cite{Kozhberov22}, where Eq.\ (\ref{res}) were obtained as a result of numerical calculations for bcc and fcc lattices and to anonymous referee for useful
comments.
This research was supported by The Ministry of Science and Higher
Education of the Russian Federation (Agreement with Joint Institute for
High Temperatures RAS No 075-15-2020-785 dated September 23, 2020). 
	
	\section*{DATA AVAILABILITY}
	The data underlying this letter are available in the letter.

\bsp	
\label{lastpage}
\end{document}